\documentclass[sigconf]{acmart}
\AtBeginDocument{%
  }

\usepackage{multirow}
\usepackage{makecell}
\usepackage{algorithm}
\usepackage{algorithmic}
\usepackage{graphicx}
\usepackage{subcaption}
\usepackage{multirow}
\usepackage[normalem]{ulem}
\usepackage{float}
\usepackage{amsthm}
\usepackage{enumitem}
\newtheorem{assumption}{Assumption}

\begin{document}

\title{Uncertainty as Remedy: Mitigating Satisfaction Label Bias in Short Video Multi-Objective Ensemble Ranking}

\author{Zonghe Shao}
\authornote{These authors contributed equally to this work.}
\affiliation{
  \institution{Kuaishou Technology}
  \city{Beijing}
  \country{China}
}
\email{zhshao@hust.edu.cn}

\author{Tiantian He}
\authornotemark[1]
\affiliation{
  \institution{Kuaishou Technology}
  \city{Beijing}
  \country{China}
}
\email{hetiantian05@kuaishou.com}

\author{Xiaoxiao Xu}
\authornotemark[1]
\authornote{The corresponding author.}
\affiliation{
  \institution{Kuaishou Technology}
  \city{Beijing}
  \country{China}
}
\email{xuxiaoxiao05@kuaishou.com}

\author{Jiaqi Yu}
\affiliation{
  \institution{Kuaishou Technology}
  \city{Beijing}
  \country{China}
}
\email{yujiaqi03@kuaishou.com}

\author{Minzhi Xie}
\affiliation{
  \institution{Kuaishou Technology}
  \city{Beijing}
  \country{China}
}
\email{xieminzhi@kuaishou.com}

\author{Jinfang Gu}
\affiliation{
  \institution{Kuaishou Technology}
  \city{Beijing}
  \country{China}
}
\email{gujinfang@kuaishou.com}

\author{Yongqi Liu}
\affiliation{
  \institution{Kuaishou Technology}
  \city{Beijing}
  \country{China}
}
\email{liuyongqi@kuaishou.com}

\author{Kaiqiao Zhan}
\affiliation{
  \institution{Kuaishou Technology}
  \city{Beijing}
  \country{China}
}
\email{zhankaiqiao@kuaishou.com}

\author{Kun Gai}
\affiliation{
  \institution{Unaffiliated}
  \city{Beijing}
  \country{China}
}
\email{gai.kun@qq.com}


\begin{abstract}
The core objective of short video recommendation is to model users' unobservable true satisfaction with recommended videos.
As the dominant industrial framework, end-to-end multi-objective ensemble ranking models are typically trained with multi-dimensional dense user behavioral signals, such as clicks and watch time.
However, these behavioral signals are partial, fragmented, and often mutually conflicting user satisfaction proxies, introducing uncertainty and label bias into satisfaction modeling.
Conventional deterministic models overlook this uncertainty, which may exacerbate satisfaction label bias and result in suboptimal convergence.
This paper proposes \textbf{UAME}, an \textbf{U}ncertainty-\textbf{A}ware end-to-end \textbf{M}ulti-objective \textbf{E}nsemble ranking framework for short video recommendation. 
UAME represents the model's prediction as a Gaussian scoring variable, where the mean denotes the predicted satisfaction score and the variance quantifies predictive uncertainty associated with this score.
We further design a probabilistic pairwise ranking loss, and construct an uncertainty-aware sample-level weighting scheme to mitigate the bias. 
We further provide theoretical analysis suggesting that the weighting scheme helps mitigate satisfaction label bias.
Extensive offline and online experiments on a large-scale industrial short video platform demonstrate that UAME consistently improves two state-of-the-art paradigms, EMER and EASQ, and better aligns with questionnaire-based user satisfaction. 
UAME has been deployed in our production short-video recommendation system and continues to deliver stable, statistically significant gains.
\end{abstract}





\maketitle

\section{Introduction}

\begin{figure*}[htbp]
    \centering
    \includegraphics[width=0.8\linewidth]{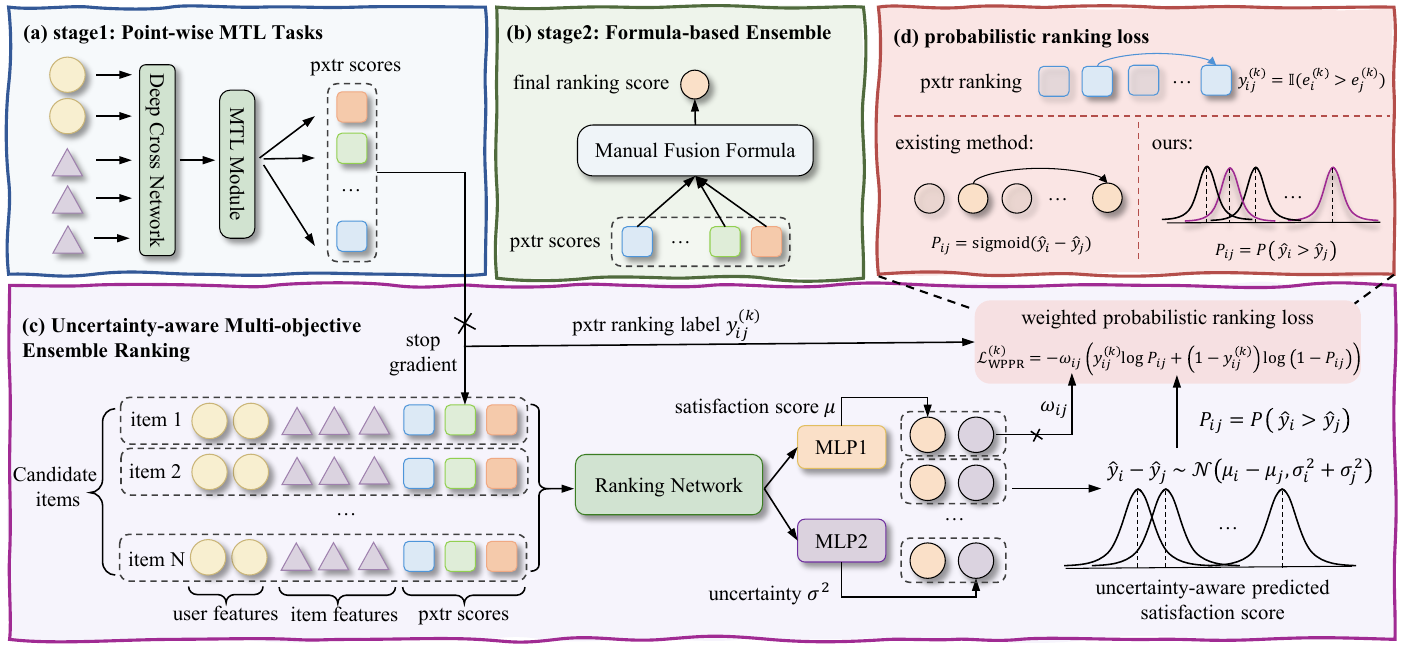}
    \caption{
    Overview of multi-objective ensemble ranking.
  (a) Stage 1: Point-wise MTL module generates pxtr scores for different user behaviors.
  (b) Stage 2: Conventional manual formula-based ensemble of pxtr scores.
  (c) The proposed uncertainty-aware multi-objective ensemble ranking framework.
  (d) Comparison of deterministic and our probabilistic pairwise ranking loss.
  }
    \label{fig:main}
\end{figure*}

Short-video recommendation platforms serve hundreds of millions of daily active users (DAUs), where the core goal of the recommender system is to accurately model the user’s unobservable true satisfaction with content~\cite{siro2023understanding}.
To model this latent variable, the dominant industrial paradigm is multi-objective ranking, which typically follows a two-stage architecture~\cite{meng2025generative}, as shown in Figures~\ref{fig:main}a and~\ref{fig:main}b.
Firstly, large-scale models are employed to estimate user satisfaction in multiple dimensions, e.g., click, watch time, like, collectively referred to as pxtr scores~\cite{li2023adatt}. 
Secondly, these pxtr scores are fused into a unified ranking score. Early approaches rely on manually designed heuristic formulas, e.g., weighted sum or product. Subsequent works have improved fusion by learning the combination weights via various methods, including cross-entropy methods~\cite{mahapatra2023multi, chen2023controllable}, grid search~\cite{wu2021multi}, and reinforcement learning~\cite{stamenkovic2022choosing, zhang2022multi}, yet they still lack end-to-end personalized ranking.

To address these bottlenecks, end-to-end multi-objective ensemble ranking~\cite{xu2025umre, he2025end, cao2025pantheon, xia2026harmonrank, li2026towards} has emerged as a promising direction. 
This paradigm replaces manually designed heuristic fusion formula with deep learning models, and uses multiple pxtr signals as user satisfaction proxies, enabling end-to-end ranking. 
Representative works such as EMER~\cite{he2025end}, Pantheon~\cite{cao2025pantheon}, and HarmonRank~\cite{xia2026harmonrank} have validated the effectiveness of this paradigm in large-scale industrial short-video recommendation scenarios, and achieved substantial performance gains over heuristic formulas.
Despite the progress, they still exhibit two important limitations that prevent alignment with users' real comprehensive satisfaction.

On the one hand, they fail to resolve the inherent bias between multi-dimensional satisfaction proxies and true user satisfaction~\cite{li2026towards}.
Existing methods implicitly treat pxtr signals as sufficiently reliable user satisfaction proxies and adopt an equal-weight loss for optimization.
In practice, heterogeneous pxtr signals only capture fragmented dimensions of short-video user behaviors~\cite{tang2024multi}, which carry ranking constraints that often contradict each other and deviate from real comprehensive preference, introducing inherent satisfaction label bias.
The equal-weight loss can further aggravate this bias, which makes items with consistent pxtr ranking dominate model updates, while high-conflict items critical to satisfaction modeling are severely under-optimized. 
On the other hand, the deterministic modeling paradigm cannot reconcile conflicting optimization objectives. 
Existing models output a deterministic score, assuming equal confidence in all predictions and that a fixed score can satisfy all conflicting pxtrs. 
Both assumptions are often violated in real-world~\cite{christakopoulou2020deconfounding}, forcing the model to converge to a compromised local optimum~\cite{li2025multi} without uncertainty quantification~\cite{wang2025uncertain}.

Additionally, existing uncertainty-aware recommendation works mostly focus on point-wise prediction tasks~\cite{du2021exploration, paliwal2024predictive}, or mainly use uncertainty as an auxiliary signal for post-hoc ranking adjustment~\cite{yang2024mitigating,lyu2025uncertainty}, with few attempts to embed it into the core end-to-end optimization pipeline to mitigate label bias. 
They also rarely clarify the source of uncertainty tightly coupled with satisfaction label bias.

In this paper, we propose the \textbf{U}ncertainty-\textbf{A}ware end-to-end \textbf{M}ulti-objective \textbf{E}nsemble ranking \textbf{(UAME)} framework. 
Our core insight is that, under multi-objective ranking labels, the predictive uncertainty driven by multi-objective conflicts is empirically associated with sample-level satisfaction label bias, which reflects the bias between multi-dimensional satisfaction proxies and true user satisfaction.
Based on this, we construct an adaptive weighting mechanism solely via predicted uncertainty, which can be embedded into the core optimization pipeline to mitigate satisfaction label bias and align the model with true user satisfaction.

The core contributions of this paper are summarized as follows:
\begin{itemize}
\item We introduce an uncertainty-aware scoring formulation for multi-objective ensemble ranking, where the mean estimates the satisfaction score and the variance captures multi-objective conflict-driven predictive uncertainty.
\item We develop a probabilistic pairwise ranking objective with uncertainty-aware sample weighting, which leverages learned uncertainty to emphasize high-conflict item pairs and mitigate satisfaction label bias during training.
\item We deploy and extensively validate the proposed framework on a large-scale industrial short-video recommendation system with hundreds of millions of DAUs.
Extensive offline experiments, online  A/B tests, and questionnaire-based analyses demonstrate its effectiveness, stability, and improved alignment with true user satisfaction.
\end{itemize}

\section{Related Works}

\subsection{End-to-end Multi-objective Ensemble Ranking}
End-to-end multi-objective ensemble ranking has become the mainstream research direction of industrial short-video recommendation, which replaces the manual fusion formula in the traditional two-stage paradigm with joint optimization of multi-objective pxtr signals.
Representative works such as EMER~\cite{he2025end}, Pantheon~\cite{cao2025pantheon}, and HarmonRank~\cite{xia2026harmonrank} have validated the effectiveness of this paradigm in large-scale industrial recommendation.
Despite their effectiveness in eliminating manual fusion constraints, all these methods share a core limitation that they rely heavily on user behavior-derived pxtr signals as ranking labels, thus introducing the inherent bias between multi-dimensional user satisfaction proxies and users' true comprehensive satisfaction. 
Existing works have attempted to resolve this bias with additional supervision. 
For example, EASQ~\cite{li2026towards} introduced explicit satisfaction feedback from user questionnaires as high-quality supervision to align the ranking model with user satisfaction.
Different from this line of works, our work analyzes the connection between the aforementioned bias and model prediction uncertainty, and proposes an uncertainty-aware adaptive weighting scheme for item pairs, which mitigates the label bias in the core end-to-end training pipeline without relying on additional supervision signals beyond the existing pxtr behavior labels.

\subsection{Uncertainty Estimation in RecSys}
Classical machine learning theory divides uncertainty into epistemic uncertainty and aleatoric uncertainty~\cite{kendall2017uncertainties}, and this work mainly discusses aleatoric uncertainty.
Early works extensively explored methods to predict uncertainty for point-wise tasks such as click-through rate (CTR) prediction~\cite{liu2019deep}, where uncertainty is typically attributed to label noise, e.g., accidental clicks~\cite{liu2023uncertainty,jiang2024deep} and distribution shift~\cite{li2023uncertainty,neupane2024evidential}.
These studies mainly use uncertainty to characterize prediction confidence in point-wise recommendation tasks.
However, end-to-end multi-objective ensemble ranking~\cite{he2025end}, which relies on multiple heterogeneous pxtr labels for supervision, introduces a new source of uncertainty in modeling.
While label noise and distribution shift still exist, the dominant source of uncertainty here is intrinsic conflicts between multiple behavior pxtr labels for the same user-item pair.
A closely related line of research is uncertainty-aware multi-task learning (MTL)~\cite{kendall2018multi}, yet these methods focus on task-level loss balancing for point-wise prediction tasks to alleviate global conflicts across different tasks~\cite{,wang2020m2grl,liang2025adaprl, wang2026dmgd}.
In contrast, our work targets sample-level satisfaction label bias in ranking caused by inconsistent multi-objective behavior labels, and models uncertainty to mitigate bias, which is fundamentally different from task-centric design of MTL and remains unexplored in multi-objective ensemble ranking.

\subsection{Uncertainty in Ranking}
Ranking lies at the core of recommendation system, aiming to align users' true satisfaction via optimal item lists.  
Inherent prediction uncertainty from label noise, behavior bias~\cite{yang2022can} and distribution shift~\cite{heuss2023predictive} has been addressed by existing works using uncertainty as a core ranking optimization signal.
Other works~\cite{yang2024mitigating} have alleviated the exploitation bias in learning-to-rank via an uncertainty-aware empirical Bayesian method, and~\cite{scharf2025rank} have modeled uncertainty in the top-K ranking problem based on score distribution modeling. 
However, existing research on uncertainty for ranking models still has a core paradigm limitation. 
Most existing works treat uncertainty as an auxiliary signal to  correct the predicted ranking scores~\cite{li2025recommenders,knyazev2023lightweight} and perform post-hoc ranking adjustment~\cite{lyu2025uncertainty}. 
Such post-hoc usage does not directly mitigate the label bias in the core optimization process.
Unlike these existing works, our method embeds uncertainty into the core training pipeline to directly mitigate the label bias during optimization, while satisfying the low-latency requirements of industrial applications.

\section{Method}
In this section, we present the core framework of \textbf{UAME}, including satisfaction-uncertainty modeling, probabilistic pairwise ranking loss, and uncertainty-aware adaptive weighting mechanism, followed by theoretical analysis.

\subsection{Problem Statement}
We formally define the end-to-end multi-objective ensemble ranking task as follows.
Let \(U\) denote the user set, \(V\) the short-video item set, and \(c\) the context features.
For a user \(u \in U\) under context \(c\), the model takes the candidate set \(I_{cand}(u,c)\) as input, and aims to generate a ranked list that maximizes the user’s unobservable true satisfaction \(s_i\) for each candidate item \(v_i\). 
The critical challenge of this task is that \(s_i\) is a latent variable unavailable for direct model supervision.
Therefore, existing methods rely on behavior-based pxtr signals as multi-dimensional user satisfaction proxies, which are generated by pre-trained MTL models for core interaction objectives such as click, watch time, like, comment, and follow.
For item \(v_i\), its \(K\)-dimensional pxtr scores are denoted as \(\{ e_i^{(k)} \}_{k=1}^K\).
For any item pair \((v_i, v_j)\), the ideal pairwise ranking label is \(y_{ij}^* = \mathbb{I}\{s_i > s_j\}\), 
and the pairwise ranking label derived from the \(k\)-th pxtr signal is \(y_{ij}^{(k)} = \mathbb{I}\{ e_i^{(k)} > e_j^{(k)} \}\), where \(\mathbb{I}\{\cdot\}\) is the indicator function.
Notably, pxtr signals are inherently biased proxies for user satisfaction, and sample-level conflicts arise across heterogeneous pxtr signals, which are the dominant sources of prediction uncertainty in end-to-end multi-objective ensemble ranking.

\subsection{Satisfaction-Uncertainty Modeling}
Conventional end-to-end multi-objective ensemble ranking methods output only a single scalar ranking score per candidate item, without quantifying prediction uncertainty. 
To address this limitation, we design an uncertainty-aware ranking architecture with a weight-sharing two-branch output layer.

As illustrated in Figure~\ref{fig:main}c, the model takes user features, item features and pxtr scores as input, which are fed into the backbone ranking network.
The two output branches fully share the refined feature representation from the backbone ranking network, and generate their respective outputs via two independent lightweight two-layer MLPs.
The score prediction branch outputs the model’s predicted satisfaction score $\mu_i$, while the uncertainty prediction branch outputs the prediction uncertainty $\sigma_i^2$.
Based on the two outputs above, we model the uncertainty-aware scoring variable $\hat{y}_i$ for the $i$-th candidate item $v_i$ as:
\begin{equation}
\label{eq normal distribution}
    \hat{y}_i = \mu_i + \sigma_i \epsilon_i,\quad \epsilon_i \sim \mathcal{N}(0,1)
\end{equation}
where $\mu_i$ denotes the predicted satisfaction score, and $\sigma_i^2$ controls the prediction uncertainty associated with this score.

\subsection{Probabilistic Pairwise Loss}

\subsubsection{Probabilistic Ranking}
For a candidate item pair \((v_i, v_j)\), the core goal of the ranking task is to model the probability that \(v_i\) should be displayed prior to \(v_j\).
To this end, we define the pairwise ranking probability $P_{i,j}$ as
\begin{equation}
    P_{i,j} = P(\hat{y}_i > \hat{y}_j)
\end{equation}
where \(\hat{y}_i\) and \(\hat{y}_j\) denote the uncertainty-aware scoring variables of \(v_i\) and \(v_j\), respectively.

For tractability, we assume that \(\hat{y}_i\) and \(\hat{y}_j\) are conditionally independent.
Based on equation~(\ref{eq normal distribution}), the pairwise score difference \(z = \hat{y}_i - \hat{y}_j\) follows $z \sim \mathcal{N}(\mu_i - \mu_j, \sigma_i^2 + \sigma_j^2)$.
From this, we derive the analytical form of the probability that \(v_i\) should be displayed prior to \(v_j\), which is the cumulative distribution function (CDF) of the standard Gaussian distribution:
\begin{equation}
    \label{eq standard normal distribution}
    P(\hat{y}_i > \hat{y}_j) = P(z > 0) = \Phi ( \frac{\mu_i - \mu_j}{\sqrt{\sigma_i^2 + \sigma_j^2}} )
\end{equation}
where \(\Phi(\cdot)\) is the CDF of the standard Gaussian distribution.

\subsubsection{Probabilistic Pairwise Ranking Loss}

Based on the ranking relation in equation (\ref{eq standard normal distribution}), we construct a probabilistic pairwise ranking (PPR) loss  to accommodate joint supervision from pxtr signals. 
For the set of all valid item pairs \(\mathcal{D}\), the PPR loss for a pxtr signal is defined as:
\begin{equation}
\label{eq loss xtr}
\begin{aligned}
\mathcal{L}_{\text{PPR}} = - \sum_{(i,j) \in \mathcal{D}} \left[ \right.
&y_{ij}^{(k)} \cdot \log\left( P(\hat{y}_i > \hat{y}_j) \right) \\
&+ \left. (1 - y_{ij}^{(k)}) \cdot \log\left( 1 - P(\hat{y}_i > \hat{y}_j) \right) \right]
\end{aligned}
\end{equation}
where \(y_{ij}^{(k)} \in \{0,1\}\) is the pairwise ranking label of the \(k\)-th pxtr.
\(y_{ij}^{(k)} = 1\) if \(v_i\) ranks prior than \(v_j\), and 0 otherwise.

The loss in equation (\ref{eq loss xtr}) enables the optimization of the satisfaction score and uncertainty.
When the model correctly predicts the ranking of an item pair (i.e., \(\mu_i > \mu_j\) with \(y_{ij}^{(k)}=1\)), the model tends to increase the mean gap $\mu_i-\mu_j$ and decrease \(\sigma_i^2\) and \(\sigma_j^2\), thereby increasing \(\frac{\mu_i - \mu_j}{\sqrt{\sigma_i^2 + \sigma_j^2}}\).
This drives \(P(\hat{y}_i > \hat{y}_j)\) closer to 1, reduces the loss, and enhances the confidence of the correct ranking decision.
Conversely, the model tends to adjust the means toward the correct order, while increasing \(\sigma_i^2\) and \(\sigma_j^2\) to reduce the absolute value of \(\frac{\mu_i - \mu_j}{\sqrt{\sigma_i^2 + \sigma_j^2}}\), masking the prediction more uncertain.

However, this mechanism introduces a potential risk that the model may take an undesired shortcut by increasing the uncertainty \(\sigma^2_{i}\) infinitely for all items to avoid learning the correct ranking of \(\mu_i\), which leads to a degenerate behavior.
To prevent this, we introduce an uncertainty regularization term~\cite{kendall2018multi} that penalizes excessively large uncertainty, guiding the model to maintain a reasonable and discriminative uncertainty distribution:
\begin{equation}
\label{eq loss reg}
\mathcal{L}_{\text{reg}} = \sum_{(i,j) \in \mathcal{D}} \log (1+\sigma^2_{i} + \sigma^2_{j})
\end{equation}

\subsubsection{Auxiliary Constraint Loss}

Although the PPR loss can jointly optimize the predicted satisfaction score and its uncertainty, the learned uncertainty is only implicitly supervised and may not explicitly reflect pxtr conflicts.
To make the learned uncertainty more explicitly characterize the level of conflict across pxtr signals, we introduce an auxiliary constraint loss, which enforces alignment between the model's uncertainty output and the level of pxtrs conflict of items.
Specifically, for each candidate item, we quantify its level of conflict as the standard deviation of its ranking positions across multiple pxtr signals. 
Based on this measure, we construct a pairwise training set $\mathcal{B}_{aux}^+$, which consists exclusively of item pairs $(i,j)$ where item $v_i$ exhibits significantly higher conflict than item $v_j$.
We adopt a pairwise logistic loss to construct the auxiliary constraint loss:
\begin{equation}
\label{eq auxiliary constraint loss}
\begin{aligned}
    \mathcal{L}_{\text{aux}} &= - \sum_{(i,j) \in \mathcal{B}_{aux}^+} \log \left( P(\sigma_i^2 \triangleright \sigma_j^2) \right)\\
    &= - \sum_{(i,j) \in \mathcal{B}_{aux}^+} \log \left( \text{sigmoid}(\sigma_i^2 - \sigma_j^2) \right)
\end{aligned}
\end{equation}
where $P(\sigma_i^2 \triangleright \sigma_j^2)$ denotes the probability that the uncertainty of item $v_i$ is higher than that of item $v_j$.

\subsection{Uncertainty-aware Adaptive Weighting}

\subsubsection{Pairwise Comprehensive Uncertainty}
In end-to-end multi-objective ensemble ranking, item pairs are not equally informative. 
Item pairs with consistent rankings across all pxtr signals are relatively easy to optimize, as their relative order is largely unambiguous. 
In contrast, item pairs with contradictory rankings across pxtrs are more informative for satisfaction modeling, since their relative order directly affects the final ranking quality.
However, conventional pairwise ranking does not capture this heterogeneity, as it assigns equal weights to all item pairs. This may over-emphasize easy pairs while under-emphasizing pairs that are more important for multi-objective ranking.

\begin{figure}[htbp]
    \centering
    \includegraphics[width=\linewidth]{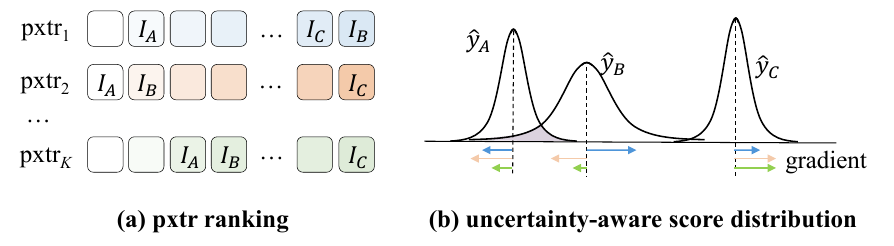}
    \caption{Multi-objective pxtrs ranking conflict and corresponding uncertainty-aware score distribution. (a) Ranking inconsistency of different candidate items under heterogeneous pxtrs. (b) Predicted score distributions with conflict-driven uncertainty.}
    \label{fig:fig2}
\end{figure}

Figure~\ref{fig:fig2}a shows the ranking inconsistency of different candidate items across heterogeneous pxtrs, while Figure~\ref{fig:fig2}b visualizes the corresponding optimization gradients and predicted satisfaction distributions.
As illustrated in Figure~\ref{fig:fig2}, under multi-objective ensemble ranking, the predicted uncertainty naturally differentiates to match this heterogeneity. 
Items with consistent pxtr rankings such as $I_A$ and $I_C$ exhibit aligned gradients, leading to low-uncertainty and  sharp distributions.
Items with severe ranking conflicts like $I_B$ suffer from contradictory gradients, leading to high-uncertainty  and broad distributions.

Leveraging this property, we quantify the level of pxtrs conflict of each item directly via its predicted uncertainty.
For an item pair \((v_i, v_j)\), we further define the pairwise comprehensive uncertainty as the sum of the predicted uncertainty of the two items:
\begin{equation}
    \label{eq comprehensive uncertainty}
    U_{i,j} = \sigma_i^2 + \sigma_j^2
\end{equation}

\subsubsection{Adaptive Weighting}
To avoid training instability caused by weight distribution, we first apply max-min normalization to the comprehensive uncertainty of all item pairs in the current candidate items, mapping pairwise uncertainty values to $[0, 1]$. 
We then scale the normalized weights by $\gamma$, which aims to ensure that high-uncertainty pairs receive appropriately enhanced weights while maintaining stable convergence of multi-objective joint training.
The adaptive weight \(\omega_{ij}\) for item pair \((v_i, v_j)\) is calculated as:
\begin{equation}
    \omega_{ij} = \gamma \times \frac{U_{i,j} - \text{min}(U)}{\text{max}(U) - \text{min}(U)}
\end{equation}
where \(\gamma\) is the scaling factor, \(\text{min}(U)\) and \(\text{max}(U)\) denote the minimum and maximum comprehensive uncertainty of all item pairs in the current training batch, respectively. 

Building on the adaptive weighting scheme derived above, we integrate the sample-level weight $\omega_{ij}$ into the PPR loss, and further combine the auxiliary constraint loss and uncertainty regularization term to form the final loss:
\begin{equation}
\label{eq loss final}
    \mathcal{L}_{\text{final}} = \sum_{k=1}^K \mathcal{L}_{\text{WPPR}}^{(k)}+ \alpha \cdot \mathcal{L}_{\text{reg}} + \beta \cdot \mathcal{L}_{\text{aux}}
\end{equation}
where $\mathcal{L}_{\text{WPPR}}^{(k)}$ is the weighted probabilistic pairwise ranking loss (WPPR) for the $k$-th pxtr signal, with the form:
\begin{equation}
\begin{aligned}
    \mathcal{L}_{\text{WPPR}}^{(k)} = - \sum_{(i,j) \in \mathcal{D}} &\omega_{ij} \cdot\left[ y_{ij}^{(k)} \cdot \log\left( P(\hat{y}_i > \hat{y}_j) \right) +\right.\\
    &\left.(1 - y_{ij}^{(k)}) \cdot \log\left( 1 - P(\hat{y}_i > \hat{y}_j) \right) \right]
\end{aligned}
\end{equation}
where $K$ is the number of pxtr signals, $\alpha > 0$ is a hyperparameter that regulates the regularization intensity of the uncertainty term, and $\beta > 0$ is a hyperparameter that balances the optimization trade-off between the weighted ranking loss and the auxiliary constraint loss.

\subsection{Theoretical Analysis}
In this subsection, we provide the theoretical analysis for uncertainty-aware adaptive weighting mechanism. 
We first characterize the satisfaction label bias in terms of ranking risk and proxy bias. 
We then establish the connection among sample-level label bias, pxtrs conflict, and uncertainty. 
Based on this result, we show that using uncertainty as the adaptive sample weight is a well-motivated way to better align the weighted proxy risk with the true ranking risk.

\subsubsection{Ranking Risk Decomposition}
The ultimate goal of the multi-objective ensemble ranking model is to model the users' true satisfaction $s$, i.e., minimize the true population risk over the real-world item pair distribution $\mathcal{P}$, defined as:
\begin{equation}
\label{eq rue population risk}
    R(f) = \mathbb{E}_{(i,j) \sim \mathcal{P}} \left[ \ell\left(P_{ij}, y^*_{ij}\right) \right]
\end{equation}
where $\ell(\cdot)$ is the pairwise ranking loss.

Existing end-to-end multi-objective ensemble ranking methods only optimize the proxy risk $R_{proxy}(f)$, which is the average loss over $K$ pxtr labels:
\begin{equation}
\label{eq proxy risk}
    R_{proxy}(f) = \mathbb{E}_{(i,j) \sim \mathcal{P}} \left[ \hat{\ell}_{ij} \right], \quad \hat{\ell}_{ij} = \frac{1}{K} \sum_{k=1}^K \ell\left(P_{ij}, y^{(k)}_{ij}\right)
\end{equation}

By the linearity of expectation, the true population risk can be formally decomposed into the following terms:
\begin{equation}
\label{eq risk decomposed}
    R(f) = R_{proxy}(f) + \Delta_{proxy}
\end{equation}
where $\Delta_{proxy} = \mathbb{E}_{(i,j) \sim \mathcal{P}} \left[ \ell\left(P_{ij}, y^*_{ij}\right) - \hat{\ell}_{ij} \right]$ is the overall inherent user satisfaction label bias, i.e.,  proxy bias.

Equation (\ref{eq risk decomposed}) reveals the core limitation of existing methods that they minimize the proxy risk $R_{proxy}(f)$, but ignore the heterogeneity of sample-level proxy bias $\Delta_{ij} = \ell\left(P_{ij}, y^*_{ij}\right) - \hat{\ell}_{ij}$. 
High-bias items with conflicting pxtr ranking labels can disproportionately affect the optimization outcome, but are assigned equal weights with low-bias items, leading to misalignment with true user satisfaction.

\subsubsection{From Proxy Bias to Predictive Uncertainty}
To establish the link between sample-level proxy bias and predicted uncertainty, we first analyze how proxy bias arises from the conflict among heterogeneous pxtr signals.

\begin{assumption}
\label{assumption:Lipschitz}
The loss $\ell(p, y)$ is $L$-Lipschitz continuous with respect to soft label $y$ when $p \in [\varepsilon, 1-\varepsilon]$, where $\varepsilon$ is a small constant.
\end{assumption}

\begin{lemma}
\label{lemma:Lipschitz}
The sample-level proxy bias satisfies
\begin{equation}
    |\Delta_{ij}| = \left|\log \frac{1-P_{ij}}{P_{ij}}\right| \cdot \left|y_{ij}^{*} - \frac{1}{K} \sum_{k=1}^K y_{ij}^{(k)} \right| \leq L \cdot \left|y_{ij}^{*} - \frac{1}{K} \sum_{k=1}^K y_{ij}^{(k)} \right|
\end{equation}
where $L=\left|\log \frac{1-\varepsilon}{\varepsilon}\right|$.
\end{lemma}

Lemma~\ref{lemma:Lipschitz} shows that the sample-level proxy bias is directly proportional to the mismatch between the true satisfaction label and the averaged pxtr labels. 
To further connect pxtrs conflict with uncertainty, we introduce the following empirically motivated assumption.

\begin{assumption}
\label{assumption:Uncertainty}
Under the optimization of the PPR loss in equation (\ref{eq loss xtr}), for an item pair $(v_i, v_j)$, $U_{ij}$ is positively correlated, in expectation, with its level of pxtrs conflict.
\end{assumption}

This assumption is consistent with the optimization behavior of PPR.
Low-conflict pairs receive consistent supervision and can be fitted with a stable mean difference and small variance, whereas high-conflict pairs impose contradictory ranking constraints and therefore tend to induce larger uncertainty. Hence,  we represent Proposition~\ref{proposition: Proxy Bias-Uncertainty}

\begin{proposition}
\label{proposition: Proxy Bias-Uncertainty}
Under Assumptions~\ref{assumption:Lipschitz}-\ref{assumption:Uncertainty}, the sample-level proxy bias $|\Delta_{ij}|$ and the pairwise comprehensive uncertainty $U_{ij}$ are positively associated in expectation through their common dependence on pxtr conflict.
\end{proposition}

\subsubsection{Rationale of Uncertainty-Aware Adaptive Weighting}
From the perspective of weighted empirical risk minimization (WERM), we define the weighted proxy risk as
\begin{equation}
\label{eq weighted empirical risk}
    R_{w}(f) = \frac{1}{|\mathcal{D}|} \sum_{(i,j) \in \mathcal{D}} \omega_{ij} \cdot \hat{\ell}_{ij}(f)
\end{equation}

Proposition~\ref{proposition: Proxy Bias-Uncertainty} suggests that the uncertainty $U_{ij}$ is positively associated, in expectation, with the magnitude of proxy bias $|\Delta_{ij}|$. 
Therefore, using uncertainty as the sample weight encourages WERM to place more emphasis on item pairs that are more likely to incur larger bias~\cite{shalev2014understanding,natarajan2013learning}.
In this sense, uncertainty-aware weighting focuses optimization on high-bias pairs and may improve alignment with the true ranking risk compared to uniform weighting.

\section{Experiments}
In this section, we conduct extensive offline and online experiments to verify the effectiveness of our proposed method, and answer the following 4 research questions (RQs):

\begin{itemize}
\item \textbf{RQ1:} Can UAME consistently improve multi-objective ranking across different backbones and pxtr objectives?
\item \textbf{RQ2:} Can UAME generalize to online user satisfaction-related metrics in the production environment?
\item \textbf{RQ3:} How do the uncertainty-related components and hyperparameters influence the effectiveness of UAME?
\item \textbf{RQ4:} Does learned uncertainty reflect satisfaction label bias and improve alignment with user satisfaction?
\end{itemize}

\subsection{Experimental Setup}
To systematically verify the effectiveness and superiority of UAME, we conduct comprehensive experiments on large-scale industrial recommendation datasets.
The detailed experimental setup is introduced as follows.

\subsubsection{Datasets}
All experiments are conducted on real-world datasets collected from a large-scale industrial short video recommendation platform with hundreds of millions of DAUs, fully aligning with real-world industrial scenarios for large-scale multi-objective recommendation systems.

\subsubsection{Evaluation Metrics}
We adopt user Group Area Under the Curve (GAUC) as the primary metric on different key pxtrs to measure the ranking performance of the model, which is more aligned with the core personalized ranking goal of recommendation systems, and can more accurately measure the model's ability to fit each user's personalized preference, compared with the global AUC metric.
GAUC is calculated as follows: 
\begin{equation}
    \text{GAUC} = \frac{\sum_{i}^{\#\text{user}} \#\text{impression}_{u_i} \cdot \text{AUC}_{u_i}}{\sum_{i} \#\text{impression}_{u_i}}
\end{equation}
where \(u_i\) is the \(i\)-th user, \(\#\text{impression}_{u_i}\) denotes the impression count, and \(\text{AUC}_{u_i}\) denotes the corresponding AUC value.

\subsubsection{Baselines}
To evaluate the effectiveness of our method on representative industrial multi-objective ensemble ranking frameworks, we select two state-of-the-art (SOTA) models as our backbone ranking networks: 
\begin{itemize}
    \item \textbf{EMER}~\cite{he2025end}, a widely deployed industrial end-to-end ensemble ranking method.
    \item \textbf{EASQ}~\cite{li2026towards}, a user satisfaction-aligned ranking method, which introduces explicit user questionnaire satisfaction feedback as high-quality ranking supervision. 
\end{itemize}

And we compare our method with four baselines:
\begin{itemize}
    \item \textbf{Base Model} refers to the EMER and EASQ backbone without our uncertainty-aware components.
    \item \textbf{RankDist}~\cite{scharf2025rank} models uncertainty by predicting variance, but only uses it to adjust ranking scores during inference.
    \item \textbf{EBRank}~\cite{yang2024mitigating} adopts empirical Bayes to model prediction uncertainty and utilizes upper confidence bounds (UCB) for exploration.
    \item \textbf{ANSL}~\cite{gao2025both} assigns optimization weights to samples automatically according to the magnitude of their ranking loss.
\end{itemize}

Together, these baselines correspond to different settings: 
using uncertainty only at inference (RankDist, EBRank) or applying sample weighting without uncertainty (ANSL), which highlights the benefit of using uncertainty as sample-level weights during training.

\subsubsection{Implementation Details}
For fair comparison, all methods are evaluated under the same data split, feature setting, and evaluation protocol. 
For each baseline, we independently tune its key hyperparameters on the validation set to obtain a strong configuration.
For our method, the embedding size of user and item features is fixed at 32, and the embedding dimension of the pxtr is fixed at 8. 
Training is performed using the Adagrad optimizer~\cite{duchi2011adaptive} on the shuffled samples, while learning rate is set to 0.001. 
The loss weights \(\alpha\) and \(\beta\) in equation (\ref{eq loss final}) are set to 0.02 and 0.1, respectively, and the weight scaling factor \(\gamma\) is set to 2.
More implementation and deployment details are provided in Appendix~\ref{app:implementation}.

\begin{table*}[t]
  \caption{Performance comparisons between UAME and the baselines across two backbones. The best and second-best performance methods are denoted in bold and underlined fonts, respectively.}
  \label{tab:Overall Performance}
  \centering
  \begin{tabular}{llcccccccc}
    \toprule
    \multicolumn{2}{c}{\textbf{Methods}} & \textbf{pctr} & \textbf{pvtr} & \textbf{plvtr} & \textbf{pcpr} & \textbf{pltr} & \textbf{pwtr} & \textbf{pcmtr} & \textbf{pftr} \\
    \midrule
    \multirow{5}{*}{\makecell{EMER~\cite{he2025end}}}  & Base Model  & \underline{0.706} & 0.667 & \underline{0.731} & 0.640 & 0.688 & 0.684 & 0.686 & \underline{0.706} \\
         & RankDist~\cite{scharf2025rank}  & 0.700 & 0.663 & 0.729 & \underline{0.649} & 0.690 & 0.677 & 0.678 & 0.689 \\
         & EBRank~\cite{yang2024mitigating} & 0.686 & 0.687 & 0.714 & 0.623 & \underline{0.696} & 0.707 & \underline{0.697} & 0.701\\
         & ANSL~\cite{gao2025both} & 0.666 & \textbf{0.715} & 0.670 & 0.642 & 0.669 & \underline{0.709} & 0.599 & 0.692 \\
         & UAME   & \textbf{0.715} & \underline{0.699} & \textbf{0.735} & \textbf{0.657} & \textbf{0.712} & \textbf{0.724} & \textbf{0.711} & \textbf{0.719} \\
    \midrule
    \multirow{5}{*}{\makecell{EASQ~\cite{li2026towards}}}  & Base Model  & 0.672 & 0.598 & 0.668 & 0.631 & 0.682 & \underline{0.661} & \textbf{0.703} & \underline{0.666} \\
         & RankDist~\cite{scharf2025rank}  & \underline{0.707} & 0.578 & \underline{0.700} & 0.673 & 0.673 & 0.656 & 0.685 & 0.650 \\
         & EBRank~\cite{yang2024mitigating} & 0.693 & 0.634 & 0.689 & \underline{0.679} & 0.675 & 0.650 & 0.669 & 0.655\\
         & ANSL~\cite{gao2025both} & 0.670 & \underline{0.636} & 0.677 & 0.632 & \textbf{0.712} & 0.666 & 0.683 & 0.656 \\
         & UAME   & \textbf{0.734} & \textbf{0.669} & \textbf{0.763} & \textbf{0.686} & \underline{0.684} & \textbf{0.675} & \underline{0.695} & \textbf{0.678} \\
    \bottomrule
  \end{tabular}
\end{table*}

\subsection{Overall Performance (\textbf{RQ1})}
To answer \textbf{RQ1}, we conduct comparative experiments on the EMER and EASQ backbones. 
Table~\ref{tab:Overall Performance} presents the offline performance across eight core industrial pxtrs, including 
\textbf{pctr}/ click, 
\textbf{pvtr}/ watch-time, 
\textbf{plvtr}/ long-view, 
\textbf{pcpr}/ complete-view, 
\textbf{pltr}/ like, 
\textbf{pwtr}/ follow, 
\textbf{pcmtr}/ comment, 
and \textbf{pftr}/ share.
From these results, we derive the following core conclusions:

(1) Our method achieves overall improvements across both EMER and EASQ backbones, and shows relatively consistent gains on a majority of pxtr metrics. 
The consistent gains across two structurally distinct backbones also demonstrate the strong generalization ability of our framework, which can be easily embedded into mainstream industrial multi-objective ensemble ranking models without modifying the core backbone structure.

(2) Our method consistently outperforms the original backbones. Compared with the original EMER model, our method brings up to 5.8\% relative improvement on core business metrics (e.g., pwtr from 0.684 to 0.724). On the EASQ backbone, the improvement is more prominent, with up to 14.2\% relative gain on plvtr (from 0.668 to 0.763) and 9.2\% on pctr (from 0.672 to 0.734).

(3) The comparison with baselines fully validates the validity of our core design. 
RankDist and EBRank, which use uncertainty for inference-stage score adjustment, only achieve partial improvement on a few metrics, suggesting that post-processing may not fully exploit the value of uncertainty. 
ANSL, which assigns weights purely by loss magnitude, causes performance drops on several key metrics, indicating that generic sample hardness is insufficient for characterizing the label bias in multi-objective ensemble ranking. 
In contrast, our method prioritizes high-uncertainty pairs via uncertainty-characterized label bias, achieving more stable overall performance without sacrificing major objectives.

\subsection{Online A/B Testing (\textbf{RQ2})}

To answer \textbf{RQ2} and validate the real-world performance of UAME, we conduct online A/B tests over 7 days on an industrial short-video platform. 
We set up two parallel test groups to verify UAME’s generalization across two representative industrial SOTA backbones including EMER and EASQ, with each group independently allocated 5\% of the main feed traffic to enable statistically powered evaluation.
The first group uses the fully deployed online EMER model as control, with the experimental group adopting EMER integrated with UAME. The second group uses the user satisfaction-aligned EASQ model as control, with the experimental group adopting EASQ integrated with UAME.

We evaluate a comprehensive set of user satisfaction metrics, covering long-term retention (LT7), implicit consumption behavior (App stay time, watch time, video view count), and explicit positive interaction (Like, follow, forward, comment). 
As summarized in Table~\ref{tab:ab_test_result}, UAME delivers statistically significant improvements across all metrics in both test groups, with the most prominent gains on longviewing and active interaction metrics. 
Although some absolute gains are numerically small, this is common in mature large-scale recommendation systems, where improvements are typically incremental but consistent across multiple key metrics.
These consistent gains indicate that UAME effectively mitigates satisfaction label bias and aligns model optimization with true user satisfaction in large-scale industrial environments.

\begin{table}[t]
  \caption{Online A/B performance comparison between UAME and the Baseline Method. The p-value is less than 0.005.}
  \label{tab:ab_test_result}
  \centering
  \begin{tabular}{lcc}
    \toprule
    \textbf{Metrics} & \textbf{vs EMER} & \textbf{vs EASQ} \\
    \midrule
    LT7 & +0.009\% & +0.015\%\\
    AppStayTime & +0.050\% & +0.074\%\\
    WatchTime & +0.030\% & +0.221\%\\
    VideoView & +0.536\% & +0.373\%\\
    LongView & +1.614\% & +1.126\%\\
    Like & +0.939\% & +1.349\%\\
    Follow & +0.641\% & +1.299\%\\
    Forward & +1.325\% & +1.598\%\\
    Comment & +0.679\% & +0.642\%\\
    \bottomrule
  \end{tabular}
\end{table}

\subsection{Hyperparameter and Ablation Analysis (\textbf{RQ3})}
\begin{figure*}[htbp]
    \begin{subfigure}{0.33\linewidth}
        \centering
        \includegraphics[width=\linewidth]{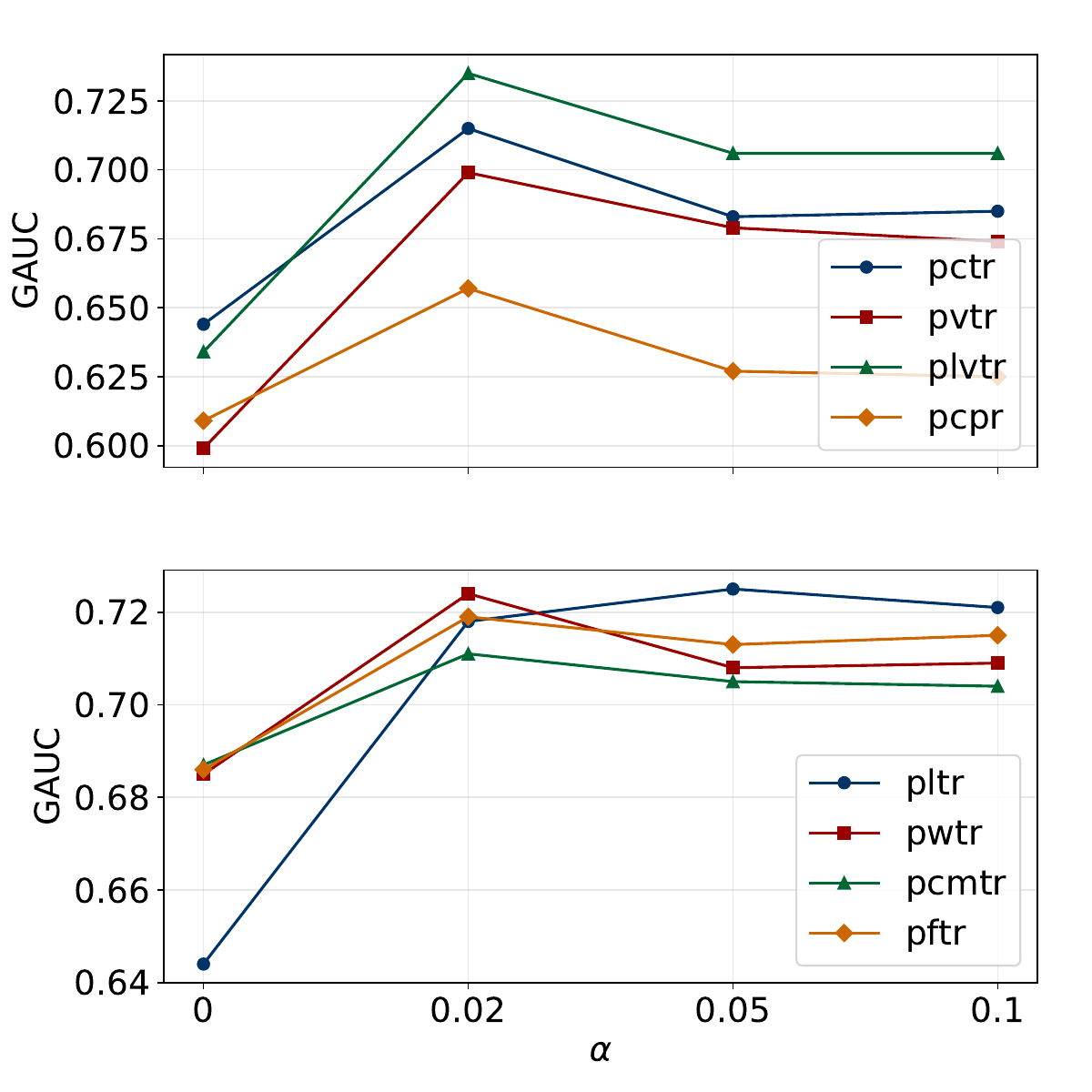}
        \caption{Effect of the regularization coefficient $\alpha$}
        \label{fig:alpha}
    \end{subfigure}
    \hfill
    \begin{subfigure}{0.33\linewidth}
        \centering
        \includegraphics[width=\linewidth]{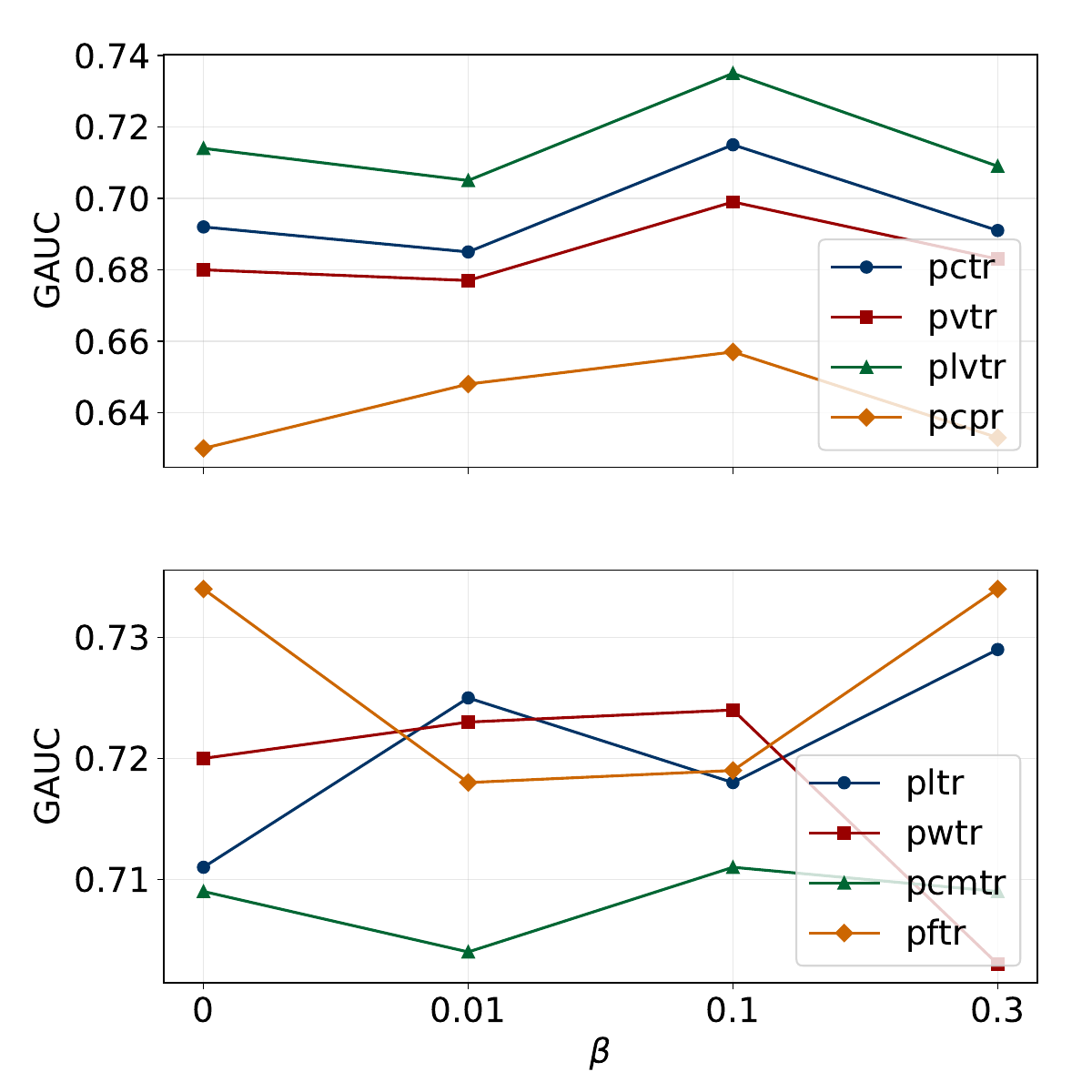}
        \caption{Effect of auxiliary loss coefficient $\beta$}
        \label{fig:beta}
    \end{subfigure}
    \hfill
    \begin{subfigure}{0.33\linewidth}
        \centering
        \includegraphics[width=\linewidth]{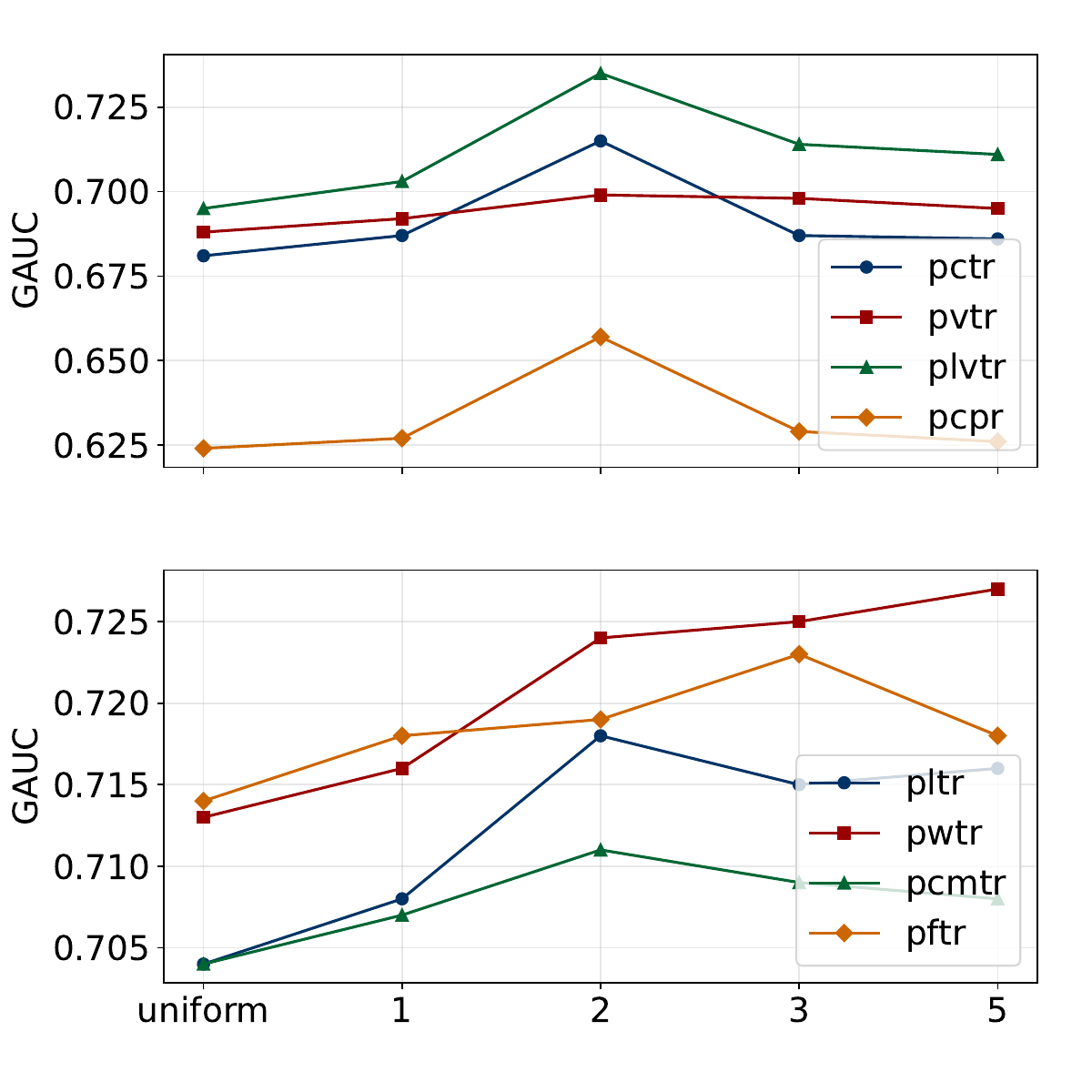}
        \caption{Effect of uncertainty scaling factor $\gamma$}
        \label{fig:gamma}
    \end{subfigure}
    \caption{Hyperparameter sensitivity analysis on EMER backbone. 
    (a) Regularization coefficient $\alpha$,
    (b) Auxiliary loss coefficient $\beta$,
    (c) Uncertainty-aware weighting and adaptive weight scaling factor $\gamma$.}
    \label{fig:hyperparameter}
\end{figure*}

To answer \textbf{RQ3} and verify the robustness of our framework, we conduct hyperparameter analysis and ablation analysis on the key components:
the regularization coefficient $\alpha$,
the auxiliary loss coefficient $\beta$, 
and the uncertainty normalization scaling factor $\gamma$.
All experiments are implemented on the EMER backbone with consistent training settings.

\subsubsection{Effect of the Regularization Coefficient $\alpha$}
As shown in Figure \ref{fig:alpha}, we test $\alpha \in \{0, 0.02, 0.05, 0.1\}$, and the performance first increases and then decreases as $\alpha$ increases. 
When $\alpha$ is set to 0.02, our model achieves the best performance. An excessively small $\alpha$ cannot regularize the uncertainty properly, while an excessively large $\alpha$ excessively suppresses the discrimination of uncertainty and weakens the adaptive weighting mechanism.

\subsubsection{Effect of the Auxiliary Loss Coefficient $\beta$}
In Figure \ref{fig:beta}, we test $\beta \in \{0, 0.01, 0.1, 0.3\}$, and the model performs optimally when $\beta=0.1$. 
A small $\beta$ leads to poor alignment between uncertainty and the level of real multi-objective pxtrs conflict, whereas a large $\beta$ makes the auxiliary loss dominate the training and impair the primary ranking task.

\subsubsection{Effect of the Uncertainty Scaling Factor $\gamma$}
Figure~\ref{fig:gamma} illustrates the effect of the scaling factor $\gamma$. The model reaches the best performance at $\gamma=2$. A small $\gamma$ limits the enhancement of high-uncertainty pairs, while an overly large $\gamma$ leads to extreme weight distribution and unstable optimization.

\subsubsection{Ablation Study} $\alpha=0$ and $\beta=0$ in Figures~\ref{fig:alpha} and \ref{fig:beta} ablate the uncertainty regularization and auxiliary constraint loss, respectively, while the \textit{uniform} setting in Figure \ref{fig:gamma} ablates uncertainty-aware weighting by replacing the adaptive weight with a uniform weight, i.e., $\omega_{ij}=1$.
Overall, removing these components leads to inferior performance, confirming that these components all contribute to the effectiveness of UAME.

\subsection{Further Analysis (\textbf{RQ4})}
To answer \textbf{RQ4}, we further analyze UAME using the questionnaire-based user satisfaction signals on EASQ~\cite{li2026towards}. 
Specifically, we study two complementary questions: 
(i) whether UAME improves alignment with user satisfaction, and (ii) whether the learned uncertainty captures the satisfaction label bias.
\begin{table}[t]
  \centering
  \caption{User satisfaction modeling performance on EASQ backbone.}
  \label{tab:satisfaction_perf}
  \begin{tabular}{lccc}
    \toprule
    \textbf{Metrics} & \textbf{EASQ} & \textbf{UAME} & \textbf{Improvement $\uparrow$}  \\
    \midrule
    NDCG@5  & 0.3753 & \textbf{0.4080} & +8.71\% \\
    NDCG@10 & 0.3872 & \textbf{0.4160} & +7.44\% \\
    HR@1    & 0.4846 & \textbf{0.5312} & +9.62\% \\
    HR@5    & 0.8192 & \textbf{0.9063} & +10.63\% \\
    HR@10   & 0.8979 & \textbf{0.9421} & +4.92\% \\
    MRR     & 0.6145 & \textbf{0.6675} & +8.62\% \\
    \bottomrule
  \end{tabular}
\end{table}

\subsubsection{True User Satisfaction Modeling Performance}
We first evaluate whether UAME improves ranking quality with respect to user satisfaction. 
Unlike behavior-derived pxtr signals, which carry inherent bias and only capture fragmented dimensions of user preference, user questionnaire feedback more directly reflects users’ holistic subjective satisfaction with recommended videos.
Using the questionnaire-based signals in EASQ as approximate ground-truth satisfaction and strictly following the experimental settings, we measure ranking performance with standard metrics including Normalized Discounted Cumulative Gain (NDCG), Hit Ratio (HR), and Mean Reciprocal Rank (MRR).

As shown in Table~\ref{tab:satisfaction_perf}, UAME consistently outperforms EASQ across all evaluation metrics, with a maximum relative improvement of 8.71\% on NDCG@5 and 10.63\% on HR@5, the core metric for top-ranking quality. 
These results indicate that UAME improves alignment with user satisfaction beyond  optimization on user satisfaction proxies alone.

\subsubsection{Relationship between Uncertainty and Satisfaction Label Bias.}
\label{sec: Relationship between Uncertainty and Satisfaction Label Bias}
We next investigate whether the learned uncertainty effectively captures the satisfaction label bias induced by multiple user satisfaction proxies.
In industrial recommendation systems, true user satisfaction is typically unobservable, making it difficult to directly assess such bias in standard settings.
However, the questionnaire feedback in EASQ provides a reliable approximation of user satisfaction, enabling us to construct a label bias measure that quantifies the inconsistency between the averaged ranking derived from pxtr signals and those based on the questionnaire-based approximate ground-truth.
We then study the relationship between the learned uncertainty and the constructed label bias across item pairs. 
Empirically, on 3 million samples, we observe a positive correlation, with a Pearson correlation coefficient of 0.65 and a Spearman correlation coefficient of 0.67.
These findings align with the intuition behind Proposition~\ref{proposition: Proxy Bias-Uncertainty} and provides empirical support for Assumption~\ref{assumption:Uncertainty}.

\section{Conclusion}
In this work, we address the inherent bias between behavior-derived satisfaction labels and true user satisfaction in end-to-end multi-objective ensemble ranking models for short-video recommendation. 
We propose UAME, an uncertainty-aware multi-objective ensemble ranking framework.
UAME represents the predicted satisfaction score with a Gaussian scoring variable, and uses the uncertainty as an indicator of the level of pxtrs conflict.
Based on this, we design an uncertainty-aware adaptive weighting mechanism.
Our analysis suggests that this weighting scheme helps mitigate the user satisfaction label bias.
Extensive experiments on a large-scale industrial short-video platform with hundreds of millions of DAUs show that UAME achieves consistent, significant improvements over baselines on two representative backbones and enhances alignment with true user satisfaction.

\newpage
\clearpage
\bibliographystyle{ACM-Reference-Format}
\bibliography{ref}

\newpage
\clearpage
\appendix

\section{Proofs of Theoretical Analysis}
\label{app:proof}

\subsection{Risk Decomposition}

Recall that the true ranking risk is defined as
\begin{equation}
    R(f)=\mathbb{E}_{(i,j)\sim \mathcal{P}}
  \left[
  \ell(P_{ij},y^*_{ij})
  \right]
\end{equation}
and the proxy risk optimized by pxtr ranking labels is
\begin{equation}
    R_{\mathrm{proxy}}(f)
  =
  \mathbb{E}_{(i,j)\sim \mathcal{P}}
  \left[
  \hat{\ell}_{ij}
  \right],
  \quad
  \hat{\ell}_{ij}
  =
  \frac{1}{K}
  \sum_{k=1}^{K}
  \ell(P_{ij},y^{(k)}_{ij})
\end{equation}

By adding and subtracting $\hat{\ell}_{ij}$ inside the expectation, we have
\begin{equation}
    \begin{aligned}
  R(f)
  &=
  \mathbb{E}_{(i,j)\sim \mathcal{P}}
  \left[
  \ell(P_{ij},y^*_{ij})
  \right] \\
  &=
  \mathbb{E}_{(i,j)\sim \mathcal{P}}
  \left[
  \hat{\ell}_{ij}
  \right]
  +
  \mathbb{E}_{(i,j)\sim \mathcal{P}}
  \left[
  \ell(P_{ij},y^*_{ij})-\hat{\ell}_{ij}
  \right] \\
  &=
  R_{\mathrm{proxy}}(f)+\Delta_{\mathrm{proxy}}
  \end{aligned}
\end{equation}
where
\begin{equation}
    \Delta_{\mathrm{proxy}}
  =
  \mathbb{E}_{(i,j)\sim \mathcal{P}}
  \left[
  \ell(P_{ij},y^*_{ij})-\hat{\ell}_{ij}
  \right]
\end{equation}

This proves the decomposition in equation (\ref{eq risk decomposed}).

\subsection{Proof of Lemma~\ref{lemma:Lipschitz}}
Let
\(
  \bar{y}_{ij}
  =
  \frac{1}{K}
  \sum_{k=1}^{K}
  y^{(k)}_{ij}
\)
denote the averaged pxtr-derived soft label. 
For the binary pairwise ranking loss used in equation~(\ref{eq loss xtr}), we have
\begin{equation}
    \ell(p,y)
  =
  -y\log p-(1-y)\log(1-p)
\end{equation}

Since this loss is affine with respect to $y$, the averaged proxy loss can be rewritten as
\begin{equation}
    \hat{\ell}_{ij}
  =
  \frac{1}{K}
  \sum_{k=1}^{K}
  \ell(P_{ij},y^{(k)}_{ij})
  =
  \ell(P_{ij},\bar{y}_{ij})
\end{equation}

Therefore, the sample-level proxy bias is
\begin{equation}
    \begin{aligned}
  \Delta_{ij}
  &=
  \ell(P_{ij},y^*_{ij})
  -
  \ell(P_{ij},\bar{y}_{ij}) \\
  &=
  -\left(y^*_{ij}-\bar{y}_{ij}\right)\log P_{ij}
  +
  \left(y^*_{ij}-\bar{y}_{ij}\right)\log(1-P_{ij}) \\
  &=
  \left(y^*_{ij}-\bar{y}_{ij}\right)
  \log\frac{1-P_{ij}}{P_{ij}}
  \end{aligned}
\end{equation}

Taking the absolute value gives
\begin{equation}
    |\Delta_{ij}|
  =
  \left|
  \log\frac{1-P_{ij}}{P_{ij}}
  \right|
  \cdot
  \left|
  y^*_{ij}
  -
  \frac{1}{K}
  \sum_{k=1}^{K}
  y^{(k)}_{ij}
  \right|
\end{equation}

Under Assumption~\ref{assumption:Lipschitz}, $P_{ij}\in[\epsilon,1-\epsilon]$, and thus
\begin{equation}
    \left|
  \log\frac{1-P_{ij}}{P_{ij}}
  \right|
  \leq
  \log\frac{1-\epsilon}{\epsilon}
  =
  L
\end{equation}
Hence,
\(
  |\Delta_{ij}|
  \leq
  L\cdot
  \left|
  y^*_{ij}
  -
  \frac{1}{K}
  \sum_{k=1}^{K}
  y^{(k)}_{ij}
  \right|
\),
which completes the proof.

\subsection{Proof Sketch of Proposition~\ref{proposition: Proxy Bias-Uncertainty}}

Let $C_{ij}$ denote the level of pxtr conflict for item pair $(v_i, v_j)$. A larger $C_{ij}$ indicates stronger disagreement among the $K$ pxtr ranking pairwise labels $\{y^{(k)}_{ij}\}_{k=1}^{K}$.

From Lemma~\ref{lemma:Lipschitz}, the sample-level proxy bias satisfies
  \begin{equation}
  |\Delta_{ij}|
  \leq
  L \cdot
  \left|
  y^*_{ij}
  -
  \frac{1}{K}
  \sum_{k=1}^{K}
  y^{(k)}_{ij}
  \right|
  \end{equation}

  From the equality form in Lemma~\ref{lemma:Lipschitz}, we have
  \begin{equation}
  |\Delta_{ij}|
  =
  \left|
  \log \frac{1-P_{ij}}{P_{ij}}
  \right|
  \cdot
  \left|
  y^*_{ij}-\bar{y}_{ij}
  \right|
  \end{equation}
  
  Since $P_{ij}\in[\epsilon,1-\epsilon]$, the coefficient
  $\left|\log\frac{1-P_{ij}}{P_{ij}}\right|$ is finite and non-negative. Therefore, the conditional expectation of $|\Delta_{ij}|$ with respect to a conflict level $C_{ij}$
  can be written as
  \begin{equation}
  \mathbb{E}\left[|\Delta_{ij}| \mid C_{ij}\right]
  =
  \mathbb{E}\left[
  \left|
  \log \frac{1-P_{ij}}{P_{ij}}
  \right|
  \cdot
  \left|
  y^*_{ij}-\bar{y}_{ij}
  \right|
  \mid C_{ij}
  \right]
  \end{equation}

For item pairs with larger pxtr conflict, the averaged proxy label $\bar{y}_{ij}$ has a larger expected deviation from the true satisfaction label. Formally, define
  \begin{equation}
  g_m(c)
  =
  \mathbb{E}\left[
  \left|
  y^*_{ij}-\bar{y}_{ij}
  \right|
  \mid C_{ij}=c
  \right]
  \end{equation}

We use the following monotonic characterization to formalize the relation between pxtr conflict and the unobservable proxy-label mismatch:
  \begin{equation}
  g_m(c_1) \leq g_m(c_2),
  \quad \forall c_1 \leq c_2
  \end{equation}

Combining this condition with the equality form of Lemma~\ref{lemma:Lipschitz}, whose coefficient is non-negative and bounded, we define
  \begin{equation}
  g_\Delta(c)
  =
  \mathbb{E}\left[
  |\Delta_{ij}|
  \mid C_{ij}=c
  \right],
  \end{equation}
This supports the following monotonic tendency:
  \begin{equation}
  g_\Delta(c_1) \leq g_\Delta(c_2),
  \quad \forall c_1 \leq c_2
  \end{equation}

On the other hand, Assumption~\ref{assumption:Uncertainty} gives the monotonic relation between pxtr conflict and pairwise uncertainty. Define
  \begin{equation}
  g_U(c)
  =
  \mathbb{E}\left[
  U_{ij}
  \mid C_{ij}=c
  \right].
  \end{equation}

Then Assumption~\ref{assumption:Uncertainty} implies
  \begin{equation}
  g_U(c_1) \leq g_U(c_2),
  \quad \forall c_1 \leq c_2
  \end{equation}

Thus, both $g_\Delta(c)$ and $g_U(c)$ are non-decreasing functions of the same conflict variable $C_{ij}$. 
This shows that the expected proxy-bias magnitude and the expected pairwise uncertainty vary in the same direction as pxtr conflict increases.
Therefore, $|\Delta_{ij}|$ and $U_{ij}$ exhibit a positive expectation-level association through their common dependence on pxtr conflict, which establishes the expectation-level association in Proposition~\ref{proposition: Proxy Bias-Uncertainty}.

\section{Implementation and Deployment Details}
\label{app:implementation}
Algorithm~\ref{alg:uame} summarizes the training and online serving procedures of UAME.
During training, UAME predicts both the satisfaction score $\mu_i$ and the predictive uncertainty $\sigma_i^2$ for each candidate item. 
The uncertainty is used to construct the probabilistic pairwise ranking loss and the uncertainty-aware adaptive weights. During online serving, only $\mu_i$ is used as the final ranking score, while $\sigma_i^2$ is not involved in ranking-score computation.
Therefore, UAME does not introduce additional ranking-stage inference latency compared with the corresponding backbone model. 
In practice, the online serving path remains the same as the original end-to-end ensemble ranking backbone, except that the trained $\mu_i$ branch is used for final scoring. 
This makes UAME compatible with existing industrial serving systems without changing the online ranking pipeline.
  
For numerical stability, the uncertainty branch outputs the log-variance $\log\sigma_i^2$ instead of directly predicting $\sigma_i^2$. The variance is then obtained by exponentiation to ensure non-negativity. 
In addition, a small constant $\epsilon$ is added to the denominator of uncertainty normalization and to probability-related computations to avoid division by zero, overflow, or undefined logarithms.

\newcommand{\algstage}[1]{%
    \STATE \noindent\colorbox{gray!15}{%
      \makebox[\dimexpr\linewidth-2\fboxsep\relax][l]{#1}%
    }%
    \vspace{-1.0\baselineskip}%
  }

\begin{algorithm}[t]
  \caption{Training and Serving of UAME}
  \label{alg:uame}
  \begin{algorithmic}[1]
  \REQUIRE Candidate set $\mathcal{I}_{cand}(u,c)$, user/context features, item features, pxtr scores $\{e_i^{(k)}\}_{k=1}^{K}$, hyperparameters $\alpha,\beta,\gamma$
  \ENSURE Trained UAME model and online ranking list

  % \STATE \textbf{Training Stage}
  \algstage{//\textit{Offline Training Stage}}
  \FOR{each training batch}
      \STATE Encode user, context, item features, and pxtr scores with the backbone ranking network.
      \STATE Predict $\mu_i$ and $\log\sigma_i^2$ for each candidate item $v_i$.
      \STATE Obtain $\sigma_i^2$ from $\log\sigma_i^2$.
      \FOR{each valid item pair $(v_i,v_j)\in\mathcal{D}$}
          \STATE Construct pxtr ranking pairwise labels $y^{(k)}_{ij}=\mathbb{I}\{e_i^{(k)}>e_j^{(k)}\}$.
          \STATE Compute $P_{ij}=\Phi\left((\mu_i-\mu_j)/\sqrt{\sigma_i^2+\sigma_j^2}\right)$.
          \STATE Compute pairwise uncertainty $U_{ij}=\sigma_i^2+\sigma_j^2$.
      \ENDFOR
      \STATE Normalize $\{U_{ij}\}$ and compute $\omega_{ij}=\gamma \cdot \frac{U_{ij}-\min(U)}{\max(U)-\min(U)+\epsilon}$.
      \STATE Compute $\mathcal{L}_{WPPR}$ by equation (\ref{eq loss xtr}).
      \STATE Compute $\mathcal{L}_{reg}$ by equation (\ref{eq loss reg}).
      \STATE Compute $\mathcal{L}_{aux}$ by equation (\ref{eq auxiliary constraint loss}).
      \STATE Optimize $\mathcal{L}_{final}=\sum_{k=1}^{K}\mathcal{L}^{(k)}_{WPPR}+\alpha\mathcal{L}_{reg}+\beta\mathcal{L}_{aux}$.
  \ENDFOR

   \algstage{//\textit{Online Serving Stage}}
  % \STATE \textbf{Online Serving Stage}
  \FOR{each online request}
      \STATE Encode candidate items with the trained backbone ranking network.
      \STATE Predict the satisfaction score $\mu_i$ for each candidate item.
      \STATE Rank candidate items by $\mu_i$ and return the final ranked list.
  \ENDFOR

  \end{algorithmic}
  \end{algorithm}

\section{Additional Analysis of Satisfaction Label Bias and Uncertainty}
\label{app:bias_uncertainty}

\begin{figure}[H]
    \centering
    \includegraphics[width=\linewidth]{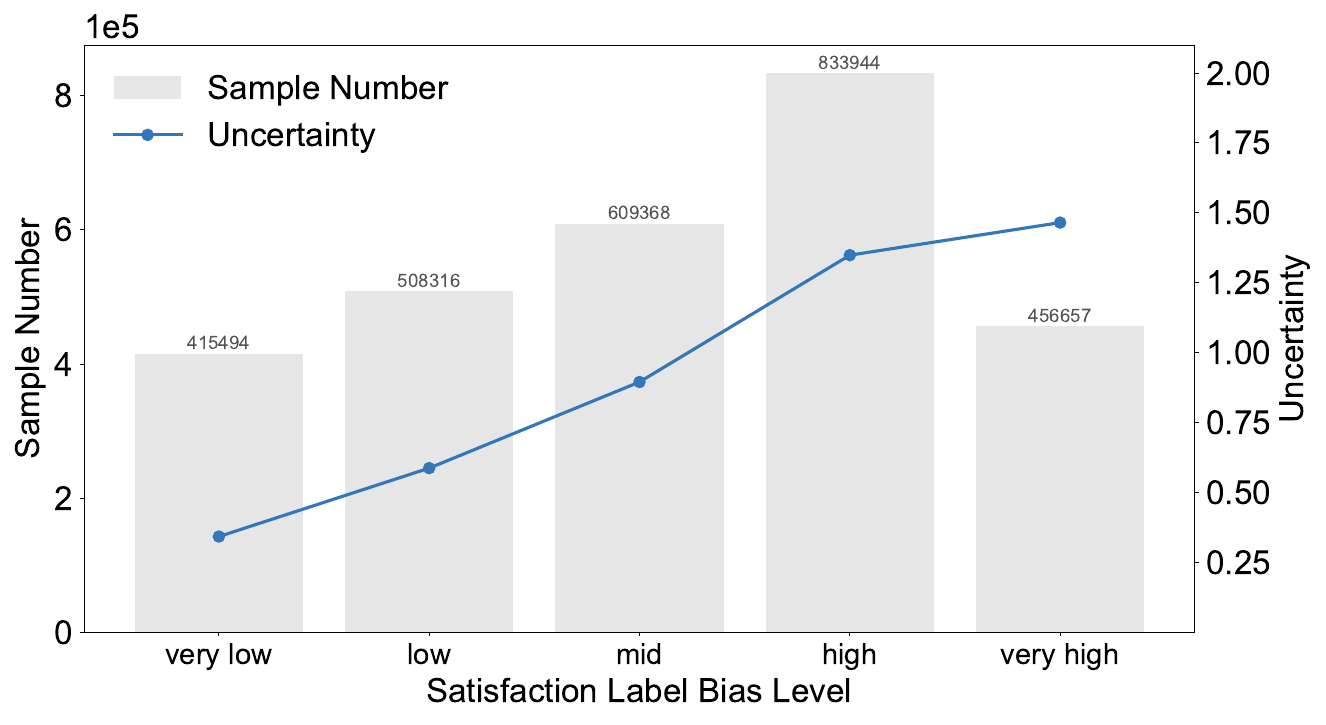}
    \caption{Bucket-level visualization of satisfaction label bias and learned uncertainty.}
    \label{fig:bias_uncertainty}
\end{figure}

To further analyze the relationship between satisfaction label bias and learned uncertainty in Section~\ref{sec: Relationship between Uncertainty and Satisfaction Label Bias}, we provide a bucket-level visualization in  Figure~\ref{fig:bias_uncertainty}.
We compute the satisfaction label bias for each item pair as $B_{ij}=\left|y^{q*}_{ij}-\frac{1}{K}\sum_{k=1}^{K}y^{(k)}_{ij}\right|$, where $y^{q^*}_{ij}$ is the questionnaire-based pairwise satisfaction label.

As shown in Figure~\ref{fig:bias_uncertainty}, item pairs are grouped into five buckets according to $B_{ij}$, from very low to very high satisfaction label bias. 
The gray bars denote the number of samples in each bucket, and the blue curve reports the mean pairwise uncertainty within each bucket. 
The uncertainty increases monotonically as the satisfaction label bias becomes larger, providing an intuitive visualization of the positive relationship between label bias and uncertainty.
This trend is consistent with the Pearson and Spearman correlations reported in Section~\ref{sec: Relationship between Uncertainty and Satisfaction Label Bias}.

\end{document}